\documentclass[prd,a4paper,superscriptaddress,nofootinbib,longbibliography]{revtex4-2}

\pdfoutput=1

\usepackage[utf8]{inputenc}
\usepackage[english]{babel}
\usepackage{mathrsfs}
\usepackage{mathtools}
\usepackage{euscript}
\usepackage{graphicx}
\usepackage{amsmath}
\usepackage{amssymb}
\usepackage{gensymb}
\usepackage{bm}
\usepackage[usenames,dvipsnames,svgnames,table]{xcolor}
\definecolor{linkcolor}{rgb}{0.6,0,0}
\definecolor{citecolor}{rgb}{0,0,0.75}
\definecolor{urlcolor}{rgb}{0.12,0.46,0.7}
\usepackage{xspace}
\usepackage{wasysym}
\usepackage{times}
\usepackage{appendix}
\usepackage{comment}
\usepackage{lipsum}
\usepackage[nolist,nohyperlinks]{acronym}
\usepackage{float}
\usepackage{natbib, ifthen}
\usepackage[breaklinks, colorlinks, urlcolor=urlcolor, linkcolor=linkcolor,citecolor=citecolor,pdfencoding=auto]{hyperref}
\usepackage{longtable}
\usepackage{listings}
\usepackage{bbm}
\usepackage{booktabs}
\usepackage{xcolor}

\definecolor{darkblue}{RGB}{0, 0, 139}  

\newcommand{\la}{\langle}
\newcommand{\ra}{\rangle}

\providecommand{\planck}{\textit{Planck}}

\newcommand{\mksym}[1]{\ifmmode {\rm #1}\else #1\fi}

\providecommand{\planck}{\Planck}

\providecommand{\lea}{\la}

\providecommand{\alt}{\lea}

\providecommand{\text}[1]{\rm{#1}}

\providecommand{\GetDist}{{\tt GetDist}}

\newcommand{\begm}{\begin{pmatrix}}
\newcommand{\enm}{\end{pmatrix}}

\newcommand\ba{\begin{eqnarray}}
\newcommand\ea{\end{eqnarray}}
\newcommand\bea{\begin{eqnarray}}
\newcommand\eea{\end{eqnarray}}

\newcommand\be{\begin{equation}}
\newcommand\ee{\end{equation}}

\newcommand{\vdelta}{\boldsymbol{\delta}}

\providecommand{\var}{\text{var}}

\newcommand{\ud}{{\rm d}}

\newcommand{\mC}{\bm{C}}

\newcommand{\mI}{\bm{I}}

\newcommand{\mM}{\bm{M}}

\newcommand{\boldvec}[1]{{\mbox{\boldmath{$#1$}}}}

\newcommand{\vX}{\boldvec{X}}

\newcommand{\vf}{\boldvec{f}}

\newcommand{\vx}{\boldvec{x}}
\newcommand{\vy}{\boldvec{y}}

\newcommand{\clo}{\mathcal{O}}

\newcommand{\Nindep}{N^{\rm indep}_{\rm eff}}
\newcommand{\Neffvar}{N^{\rm var}_{{\rm eff},X}}
\newcommand{\NeffMCMC}{N^{\rm KDE}_{{\rm eff},X}}

\newcommand{\MISE}{\text{MISE}}
\newcommand{\AMISE}{\text{AMISE}}

\newcommand{\code}[1]{%
  \ifmmode
    \text{\texttt{\textcolor{darkblue}{\detokenize{#1}}}}%
  \else
    \texttt{\textcolor{darkblue}{\detokenize{#1}}}%
  \fi
}

\begin{document}

\title{GetDist: a Python package for analysing Monte Carlo samples}

\author{Antony Lewis}
\homepage{https://cosmologist.info}
\affiliation{Department of Physics \& Astronomy, University of Sussex, Brighton BN1 9QH, UK}
\date{\today}

\begin{abstract}
Monte Carlo techniques, including MCMC and other methods, are widely used in Bayesian inference to generate sets of samples from a parameter space of interest. 
The Python \GetDist\ package provides tools for analysing these samples and calculating marginalized one- and two-dimensional densities using Kernel Density Estimation (KDE). Many Monte Carlo methods produce correlated and/or weighted samples, for example produced by MCMC, nested, or importance sampling, and there can be hard boundary priors. \GetDist's baseline method consists of applying a linear boundary kernel, and then using multiplicative bias correction.
The smoothing bandwidth is selected automatically following~\citet{botev2010}, based on a mixture of heuristics and optimization results using the expected scaling with an effective number of samples (defined here to account for both MCMC correlations and weights). Two-dimensional KDE uses an automatically-determined elliptical Gaussian kernel for correlated distributions. The package includes tools for producing a variety of publication-quality figures using a simple named-parameter interface, as well as a graphical user interface that
can be used for interactive exploration. It can also calculate convergence diagnostics, produce tables of limits, and output in LaTeX, and is publicly available.
\end{abstract}
\maketitle

\vskip .2in

\section{Introduction}

Monte Carlo (MC) techniques, including Markov Chain Monte Carlo (MCMC), nested sampling, importance sampling, and direct simulation, form the backbone of modern computational statistics and Bayesian inference~\cite{Metropolis53,Hastings70,Neal93}. 
Once samples have been generated, many (but not all) quantities of interest can easily be estimated from the samples, including parameter means, credible intervals and marginalized densities. \GetDist\ is a tool for computing these and making publication-quality figures, and is available as a Python package%
\footnote{\url{https://getdist.readthedocs.io/}, install using {\tt pip install getdist}. Source code at \url{https://github.com/cmbant/getdist/}}.  

Obtaining accurate and smooth density estimates presents particular challenges, including: (1) determining appropriate smoothing scales that balance bias and variance, (2) handling boundary effects from prior constraints, (3) accounting for correlations between samples, and (4) dealing with weighted samples. While simple histogram-based approaches are commonly used, more sophisticated methods can provide significantly more accurate results. A Bayesian approach could attempt to solve for the distribution of the true density given the samples (and a model of how they were drawn), for example using a Gaussian process prior~\cite{PrescottAdams2009b,PrescottAdams2009,Donner18}. While conceptually appealing and potentially very accurate, these solutions typically involve a further step of MC sampling and can have non-trivial computational cost. There are also practical difficulties to making it very rigorous, for example one rarely has a good model for the exact sampling distributions of realistic MCMC chains. Instead, we focus on fast and relatively simple conventional kernel density estimates (KDE), which effectively amounts to using intelligently smoothed histograms with appropriately chosen smoothing widths.

\GetDist\ provides a comprehensive solution through carefully optimized KDE implementations, alongside tools for convergence diagnostics, statistical analysis, and publication-quality visualization (using {\sc matplotlib}~\cite{Hunter:2007}). The package implements state-of-the-art bandwidth selection methods, boundary-corrected kernels, and bias reduction techniques, while remaining computationally efficient even for large sample sets. For Monte Carlo samples, one approach could be to generate a sufficiently large number of samples such that sampling noise becomes negligible. This approach could be taken when the sampling cost is low enough, allowing use of a very narrow smoothing widths. However, using a good density estimate can dramatically reduce the number of samples that are required for a given target accuracy, potentially greatly reducing the computational cost needed to produce reliable results and nice figures. 

\begin{figure}
\begin{center}
\includegraphics[width=12cm]{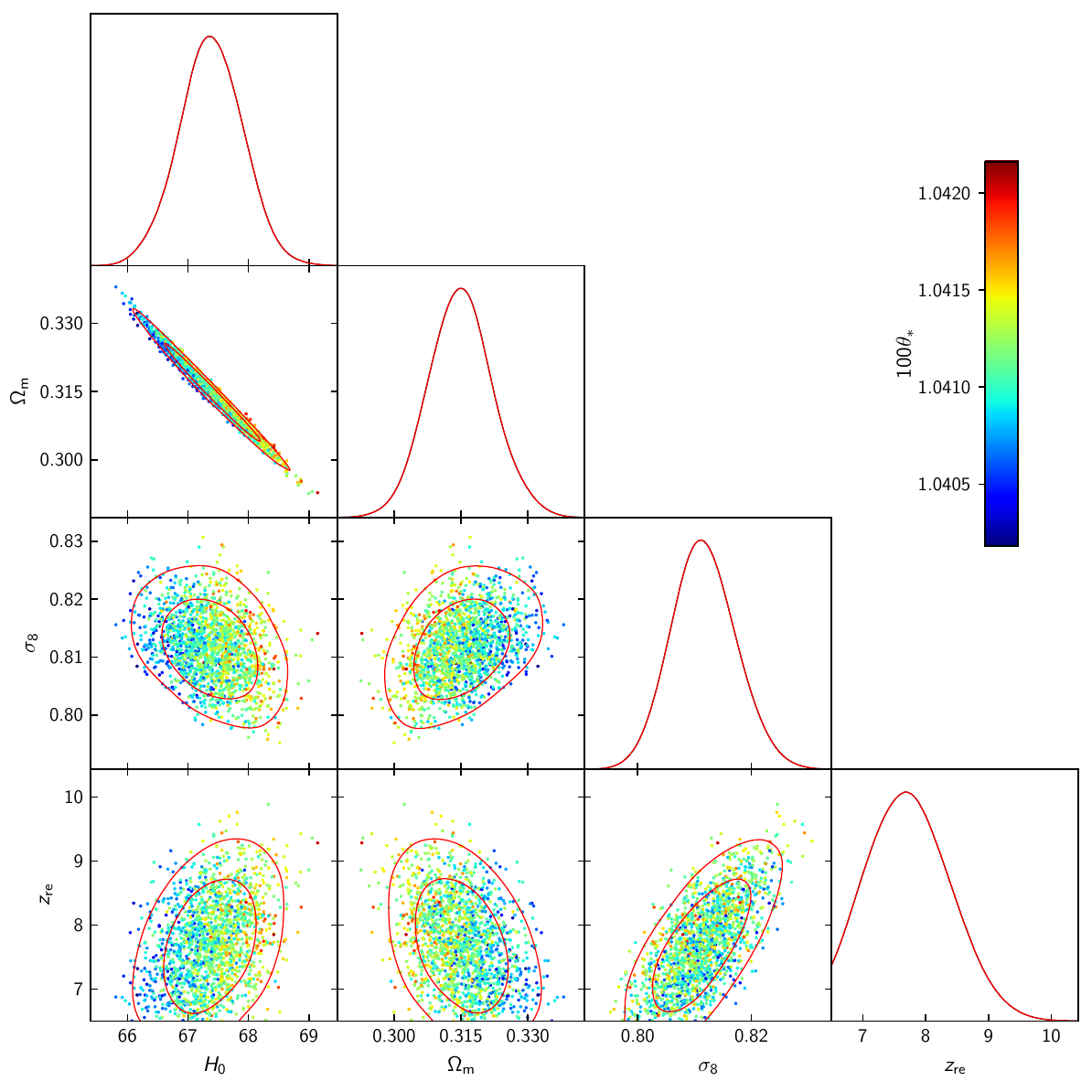}
\caption{
Example \GetDist\ `triangle' plot of MCMC parameter samples, here taken from the Planck 2018 baseline cosmological parameter chains~\cite{PCP2018} (generated using the fast-parameter dragging sampler~\cite{Lewis:2002ah,Neal04,Lewis:2013hha}).
Thinned samples are shown as coloured points, where the colour corresponds to the $\theta_*$ parameter shown in the colour bar (which is marginalized out by projection in the 1 and 2D plots). The 1D plots and 2D density contours containing 68\% and 95\% of the probability are constructed from all of the samples using kernel density estimates.
Although relatively simple unimodal distributions, all the marginalized 2D distributions are somewhat non-Gaussian. There is a hard prior on the parameter $z_{\rm re} > 6.5$ which must be accounted for in the density estimates, and $H_0$ and $\Omega_{\rm m}$ are tightly correlated. In \GetDist\ plots like this can be generated quickly with a single command using a list of input samples and a list of names of parameters to plot.
}
\label{fig:samples}
\end{center}
\end{figure}

An example of the package's capabilities is shown in Fig.\ref{fig:samples}, taken from the Planck satellite cosmological parameter analysis~\cite{PCP2018}. The figure demonstrates how the code handles both correlated samples and boundary priors in a high-dimensional parameter space, showing key 1D and 2D parameter constraints through marginalization. Since marginalization from samples simply corresponds to ignoring parameters, the marginalized densities are proportional to the local weighted sample density in the subspace of interest. The implementation achieves fast performance using FFT-based convolutions for kernel evaluation and efficient binning of samples to reduce computational scaling with sample size. For typical applications with $\clo(10^4)$ correlated samples, density estimates can be computed in a fraction of a second, making the method practical for both interactive use and large-scale analysis pipelines.
\GetDist\ works with independent single samples but also provides specific support for weighted samples and samples with substantial correlations, as typically produced by MCMC algorithms. Weighted samples naturally arise from importance sampling, nested sampling~\cite{Skilling04,Handley15,Feroz:2007kg}, and various other sampling techniques. While MCMC methods produce highly-correlated samples and can have multiple samples at a single point, once converged, the chain of samples from standard Metropolis sampling and variants should be stationary. The Monte Carlo sampling noise generally depends on both the correlations and the weights, so both must be accounted for when estimating an appropriate kernel smoothing.

\GetDist's key methodological contributions include:
\begin{itemize}
\item A robust bandwidth selection algorithm extending the fixed-point method of~\cite{botev2010} to account for both sample correlations and leading rectangular boundary effects
\item An efficient implementation of higher-order bias correction suitable for use in combination with prior boundaries
\item New estimators for effective sample sizes that approximately account for both correlations and weights in the context of KDE
\item Practical heuristics for robust and fast implementation.
\end{itemize}
Our boundary correction combines linear boundary kernels with multiplicative bias correction, providing robust performance even when posteriors intersect sharp prior boundaries. For visualization and analysis, the package provides both a programmatic interface and interactive graphical tools\footnote{\url{https://getdist.readthedocs.io/en/latest/gui.html}} (the `GetDist Gui'), making it accessible for fast exploration as well as giving fine control and reproducibility via scripts.

While Python offers several libraries for kernel density estimation, including \texttt{scipy.stats}, \texttt{seaborn}, \texttt{KDEpy}, and \texttt{scikit-learn}, these generally provide basic KDE functionality suitable for independent samples.  Although packages like \texttt{KDEpy} and \texttt{scikit-learn} offer features such as weighted samples and some boundary correction techniques, they lack specialized methods for handling correlated MCMC samples, robustly calculating density-based confidence contours, or directly addressing prior boundaries, all crucial for accurate parameter inference from Bayesian sampling.  Packages like \texttt{ArviZ} provide comprehensive tools for Bayesian analysis and visualization of MCMC results, but \GetDist\ uniquely combines state-of-the-art KDE specifically optimized for MCMC data with a strong focus on producing publication-quality figures and parameter constraints.

The \GetDist\ package is documented online (see the \href{https://getdist.readthedocs.io/}{documentation}), and
the plot gallery demonstrates use for a wide range of plotting and analysis tasks\footnote{\url{https://getdist.readthedocs.io/en/latest/plot_gallery.html}}.
This paper focuses on a technical description of the methods used, serving as a reference for what the code is doing. 
Section II develops our treatment of weighted samples and introduces the basic kernel density estimation framework. Section III presents our approach to bandwidth selection and boundary corrections, including the handling of multiplicative bias correction. Section IV addresses the specific challenges of correlated samples, developing estimators for effective sample sizes and correlation lengths. Section V presents validation tests on standard distributions and discusses computational considerations. We conclude with a discussion of limitations and potential improvements.

While the methods presented here were developed primarily for parameter estimation problems in cosmology, they are general and can be applied broadly where low-dimensional sample densities, constraints and plots are required. \GetDist\ plots and results have now appeared in hundreds of published papers across a wide range of topics, demonstrating its broad utility.

\section{Weighted samples}

We present results for weighted samples $\vX_i$ for generality, so each sample in the parameter space $\vx$ of interest is associated with a weight $w_i$ (which can be unity for unweighted samples). Estimators for the mean of a function $F(\vx)$ under the distribution $f(\vx)$ are then given by weighted sums over $n$ sample points:
\be
\hat{\bar{F}} = \frac{1}{N}\sum_{i=1}^n w_i F(\vX_i),
\ee
where $N = \sum_{i=1}^n w_i$.
Define $f_p(w_i,\vX_i)$ as the probability of getting sample point $\vX_i$ with weight $w_i$, taking points to be independent for now.
The expected value of the estimator of the mean is then
\begin{align}
\la \hat{\bar{F}}\ra  &\approx \frac{1}{N}\sum_{i=1}^n \int \ud \vX_i\,\ud w_i w_i F(\vX_i) f_p(w_i, \vX_i)
=\frac{1}{N}\sum_{i=1}^n  \int \ud \vX_i\,F(\vX_i) f_p(\vX_i) \int \ud w_i w_i f_p(w_i|\vX_i)
\nonumber\\
&=\frac{n}{N} \int \ud \vX_i\,F(\vX_i) f_p(\vX_i) \la w(\vX_i)\ra_p,
\end{align}
where here and below we neglect differences between $\la N \ra_p =n\la w\ra_p$ and $N$.
The estimator will therefore be unbiased, $\la \hat{\bar{F}}\ra_p = \bar{F}$, if
\be
\frac{\la w(\vX_i)\ra_p f_p(\vX_i)}{\la w \ra_p }  =f(\vX_i).
\label{fp}
\ee
This condition (Eq.~\ref{fp}) is satisfied in two important cases: First, for importance sampling with non-stochastic weights where $w(\vX_i) = \alpha f(\vX_i)/f_p(\vX_i)$ for any constant $\alpha$. Second, for MCMC chains where integer weights $w_i\ge 1$ count steps from rejected proposals at each point, giving $\la w(\vX_i) \ra_p \propto f(\vX_i)/f_p(\vX_i)$. Note that for MCMC, the samples are not independent, an issue we address below.

\section{Kernel Density Estimation (KDE)}

Kernel Density Estimation (KDE) provides a systematic way to estimate smooth probability distributions from discrete samples, offering significant advantages over simple histograms. While histograms can be sensitive to bin width and placement, KDE methods produce continuous estimates that better capture the underlying distribution's structure. The basic idea is to place a smooth kernel function (typically a Gaussian) at each sample point and sum these contributions to estimate the overall density. In practice there is a wide class of non-parametric methods of estimating probability densities from samples, for reviews of standard methods, see Refs.~\cite{KernelSmoothing,sheather2004,HansenNonParametrics,Zambom2012}.

In the context of MCMC analysis, we can continue sampling until we achieve sufficient sample size. For good convergence using standard criteria, this typically requires O(1000) independent-equivalent samples (and even more KDE-equivalent samples, see Sect.~\ref{sec:correlations}). 
This sample size is substantially larger than many traditional KDE applications, which often deal with smaller datasets. The larger sample size allows us to employ more sophisticated estimation methods that, while potentially unstable for small samples, perform well with larger sample sizes available here.
This section begins by reviewing some of standard definitions and estimators, discusses various complications due to boundaries and sample correlations, and then describe improved estimators using multiplicative bias correction.

The fundamental component is a density estimate $\hat{f}(x)$ of the form
\be
\hat{f}(\vx) = \frac{1}{n}\sum_{i=1}^n K_h(\vx-\vX_i) \approx \frac{1}{n} \sum_b  H_b(\vx_b) K_h(\vx-\vx_b),
\label{linearKDE}
\ee
where $\{\vX_i\}$ represents the set of sampled points, with a total of $n$ samples.
This is sometimes called the ``Parzen-Rosenblatt'' window estimator.
The kernel function $K_h$ can take various forms. By default, \GetDist\ employs (slightly truncated) zero-centered Gaussians, characterized by a width parameter $h$ (or more generally, a covariance matrix). 
This width parameter $h$ controls the kernel's broadness and consequently determines the smoothness of the estimated density function. 
For practical implementation with large sample sizes, and when focusing on low-dimensional densities, we can improve computational efficiency by binning the samples $\{\vX_i\}$ into a fine grid (with spacing much smaller than the kernel scale $K_h$). This produces bin counts $H_b(\vx_b)$ at bin centers $\vx_b$, where the total sample count is preserved ($n = \sum_b H_b$).

Also evaluating $\hat{f}$ as a (finely) binned density, we then have a simple convolution that is fast to evaluate using FFTs\footnote{Or directly if the number of points is relatively small. FFTs could also be replaced by fast gauss transforms (see e.g. \url{http://www.umiacs.umd.edu/~morariu/figtree/})}:
\begin{equation}
\hat{f} \approx \frac{1}{n} H \ast K_h.
\end{equation}
In general we have weighted samples, with each sample having a weight $w_i$, in which case
\begin{equation}
\hat{f}(\vx) = \frac{1}{N} \sum_{i=1}^n w_i K_h(\vx-\vX_i)  \approx \frac{1}{N} H \ast K_h,
\label{histotramKDE}
\end{equation}
where $N\equiv \sum_i w_i$, and $H_b$ is now the weighted sum of the samples in each bin.
In the continuum limit the histogram function is $H(\vx) = \sum_i w_i \delta(\vx-\vX_i)$, and using Eq.~\eqref{fp} we have
\begin{equation}
\frac{1}{N}\la H(\vx)\ra = \frac{1}{N}\sum_i \int \ud \vX_i \ud w_i f_p(w_i,\vX_i) w_i \delta(\vx-\vX_i)
= \frac{\la w\ra}{N}\sum_i \int \ud \vX_i f(\vX_i) \delta(\vx-\vX_i) \approx f(\vx),
\end{equation}
where we drop the $p$ subscript on the expectations $\la\ra$ where confusion should not arise.
The KDE estimator of Eq.~\eqref{histotramKDE} therefore has expectation
\bea
\la \hat{f}(\vx)\ra = [K_h \ast f](\vx),
\eea
which converges to $f(\vx)$ when $K_h$ tends to a delta function as the kernel width goes to zero ($h\rightarrow 0$).

\subsection{KDE bias and linear boundary kernels}
\label{sec:boundary}

\begin{figure}
\begin{center}
\includegraphics[width=6cm]{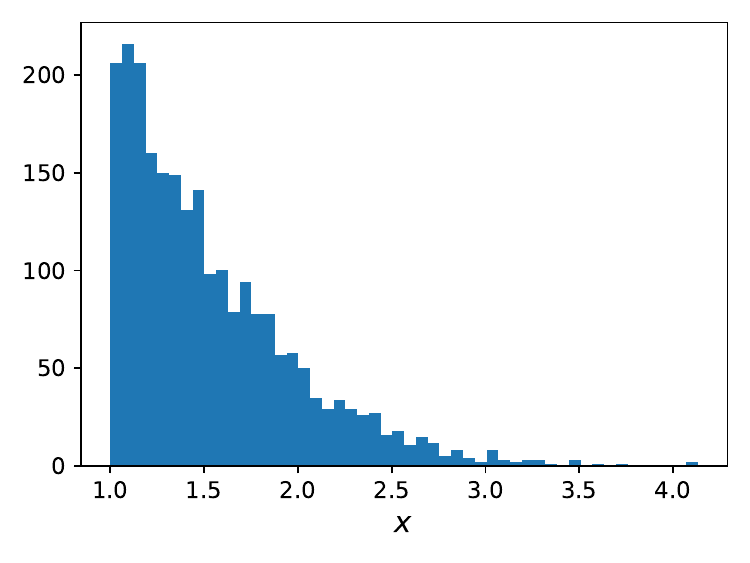}
\includegraphics[width=6cm]{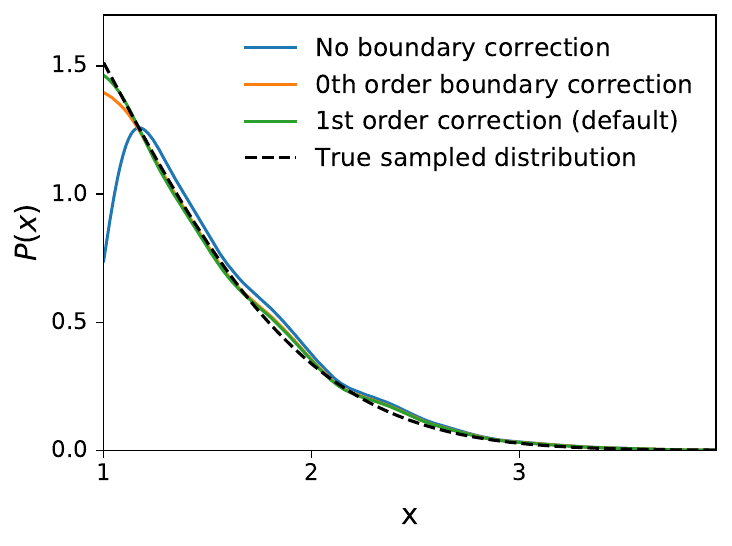}
\caption{Samples from a truncated Gaussian distribution with $x>1$. The histogram is on the left and density estimates on the right using various different kernels (without multiplicative bias correction). Some form of boundary correction is essential in order not to severely underestimate the density near the boundary. The lowest-order correction removes the leading bias, but tends to underestimate any gradient at the boundary.
In the case shown here the first-order correction works better than the zeroth order correction, but this is not guaranteed; higher-order methods can make the result less stable.
}
\label{fig:boundaries}
\end{center}
\end{figure}

Where there is a boundary, for example a prior on some parameter that it must be positive, smoothing over the boundary will give biased results, since there are no samples on one side (see Fig.~\ref{fig:boundaries}). Let's assume our function is of the form
\be
f(\vx) = B(\vx) \tilde{f}(\vx)
\ee
where $B$ is zero in the disallowed region, and one in the allowed region\footnote{
$B$ can be more general. Specifically, for the binned densities it can account for the fraction of the bin allowed by the prior (e.g. $B=1/2$ for points where the prior cuts a bin in half). It could also account for other known locations of sharp features or structure~\cite{Poluektov:2014rxa}
}, and $\tilde{f}$ is a smooth function over the scale of the kernel (and equal to $f$ where $B=1$). Series expanding $\tilde{f}$ around $\vx$ using its assumed smoothness $\tilde{f}(\vx-\vdelta) = \tilde{f}(\vx) - \tilde{\vf}^{(1)}(\vx)\cdot\vdelta+\dots$, from Eq.~\eqref{histotramKDE} in the continuum limit we have
\bea
\langle \hat{f}(\vx)\rangle &=& \frac{1}{N}
\int \langle H(\vx-\vdelta)\rangle K_h(\vdelta) d\vdelta 
= \int B(\vx - \vdelta) \tilde{f}(\vx-\vdelta) K_h(\vdelta) d\vdelta \nonumber\\
&=& \int B(\vx - \vdelta) \left[\tilde{f}(\vx)-\vdelta \cdot \tilde{\vf}^{(1)}(\vx) + \frac{1}{2} \delta^i \delta^j \tilde{f}_{ij}^{(2)}(\vx)\dots\right] K_h(\vdelta) d\vdelta \nonumber\\
&= &(K_h \ast B)\tilde{f}(\vx) - (K_h^i \ast B)  \tilde{f}^{(1)}_i(\vx)+\frac{1}{2} (K_h^{ij} \ast B)
\tilde{f}^{(2)}_{ij}(\vx) + \dots,\label{meanbias}
\eea
where $K_h^{ijk\dots}(\vx) \equiv K_h(\vx) x^i x^j x^k\dots$.
Away from the boundary where $B=1$ we have $(K_h \ast B)=1$, and
$(K_h^i \ast B)=0$ (for symmetric kernels), so the estimator is unbiased to linear order. The second order bias scales with the covariance of the
kernel ($K_h^{ij} \ast B \rightarrow [{\rm cov}(K_h)]^{ij}$) and the local curvature of $\tilde{f}$, and describes the broadening of peaks by convolution (hence typically overestimation of the variance). In units of the width of $f$, the second order bias is $\clo(h^2)$, and hence is small as long as the kernel is narrow enough compared to $f$.

\begin{figure}
\begin{center}
\includegraphics[width=6cm]{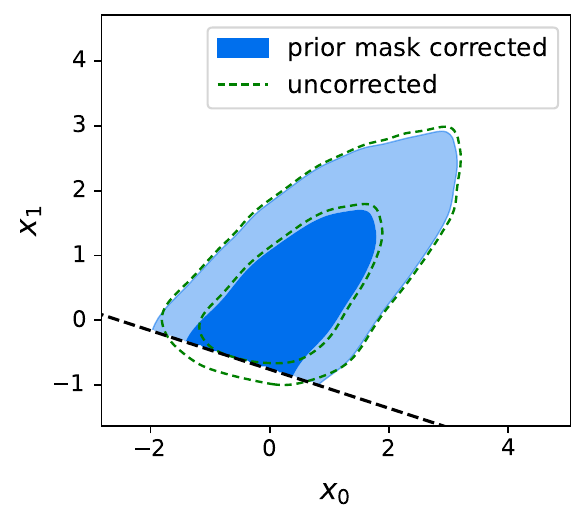}
\caption{
This figure highlights the importance of boundary correction when estimating densities with prior constraints. It shows 2D density contours for a sample distribution with a sharp linear prior constraint that is not aligned with axes (black dashed line). Without boundary correction (dashed contours), the estimated density incorrectly extends into the excluded region and the density is an artificially suppressed near the boundary due to smoothing. With boundary correction (filled contours), the density accurately reflects the true distribution and contours are correctly estimated right up to the constraint line, demonstrating the effectiveness of the boundary correction. Contours enclose 68\% and 95\% of the probability.
}
\label{fig:boundaries2D}
\end{center}
\end{figure}

With a boundary, the estimator is biased even at zeroth order. Normalizing by $(K_h \ast B)$  removes the leading bias, but leaves a linear bias if there is a non-zero gradient at the boundary. This is because the simple convolution makes the shape at the boundary too flat.
A simple solution to this is to use a linear boundary kernel~\cite{Jones93}: using a non-symmetric kernel near the boundary to remove the bias. Starting with a simple symmetric kernel $K_h$,
we can construct a more general kernel
\be
K'_h(\vx) = K_h(\vx)\left(A_0 + A_1^i x_i +\frac{1}{2} A_2^{ij}x_i x_j +\dots \right),
\ee
and solve for coefficients $\{A\}$ to make the estimator unbiased. In one dimension this is straightforward to quadratic order\footnote{Giving a fourth order kernel, see e.g. ~\cite{Jones97}}. However, it gets messy in more dimensions, and the multiplicative correction (described in Sect.~\ref{sec:multbias}) seems to be generally better at removing higher order biases. We therefore restrict to linear kernels and set $A_{\ge 2}=0$. We then have
\bea
\langle \hat{f}(\vx)\rangle &=&
 \int B(\vx - \vdelta) \tilde{f}(\vx-\vdelta)
K_h(\vdelta)\left(A_0 + A_1^i \delta_i + \dots\right) d\vdelta \nonumber\\
&=& \int B(\vx - \vdelta) \left[\tilde{f}(\vx)-\delta^i f^{(1)}_i(\vx) + \dots\right] K_h(\vdelta)\left(A_0 + A_1^j \delta_j + \dots\right) d\vdelta \nonumber\\
&= &\left[(K_h \ast B)A_0 + A_1^i (K_h^i\ast B)\right]\tilde{f}(\vx)
- \left[(K_h^i \ast B)A_0 + A_1^j (K_h^{ij}\ast B)\right] f^{(1)}_i(\vx).
\eea
Solving for unit response to $\tilde{f}$ and zero gradient bias then gives
\be
A_0 = \frac{1}{W_0-W_1^i W_2^{ij} W_1^j}
\qquad
A_1^i = -[W_2^{-1}]^{ij} W_1^jA_0,
\ee
where $W_2^{ij} \equiv (K_h^{ij}\ast B)$,
$W_1^{i} \equiv (K_h^{i}\ast B)$,
$W_0 \equiv (K_h\ast B)$. The residual bias is then $\clo(h^2)$, even approaching the boundary.
Note that the correction kernel is only different from the starting kernel within a kernel width of the boundary, since $W_0=1, W_1=0$ for symmetric kernels where $B=1$. However, for generality the terms can also be calculated by full convolutions. Using the general approach also allows incorporation of arbitrary prior boundaries that are not necessarily aligned with the parameter axes, as illustrated in Fig.~\ref{fig:boundaries2D} in a two-dimensional example.

One issue with the linear boundary kernel estimators is that they are not guaranteed to be positive.  A simple fix is to impose positivity by using the positive estimate
\be
\hat{f}_P \equiv \bar{f}\exp\left(\hat{f}/\bar{f}-1\right),
\ee
where $\bar{f}$ is the simple de-biased kernel formed by normalizing by $(K_h \ast B)$~\cite{Jones96}. We also always renormalize so that the kernel density integrates to unity (or has peak normalized to one for convenient plotting). If $\hat{f}_P$ is only used as a pilot estimate for a later higher-order estimator, accuracy of $\hat{f}_P$ near the boundary is in any case not critical.

\subsection{Statistical and total error}
To quantify the error in the kernel estimator, people often use the Mean Integrated Squared Error,
\be
\MISE \equiv \int \ud \vx \left\langle ( \hat{f}(\vx) - f(\vx))^2\right\rangle,
\label{MISE}
\ee
largely because it is convenient to calculate analytically in simple cases. There are contributions from bias and statistical noise, which trade off against each other, with broader kernels reducing the noise but increasing the bias. Assume for simplicity there are no boundary priors here, so Eq.~\eqref{meanbias} gives the leading bias
\be
\la \hat{f}\ra - f = \frac{1}{2} [\text{Cov}(K_h)]^{ij} f^{(2)}_{ij} +\dots.
\ee
To see the dependence on the smoothing scale $h$ of the  $d$-dimensional kernel, we can define the kernel as $K_h(\vx) = \frac{1}{h^d}K(\vx/h)$. We then have
\be
[\text{Cov}(K_h)]^{ij} = \frac{1}{h^d} \int \ud^d \vx x^i x^j K(\vx/h) = h^2  [\text{cov}(K)]^{ij}.
\ee
The bias is independent of whether the samples are weighted or correlated. The statistical term is more tricky however. For now, we just take the sample locations to be independent. Taking $N$ to be non-stochastic we have
\bea
\int \ud\vx \, \var \hat{f} &=& \frac{1}{N^2}\int \ud \vx \sum_i \left\{\int \ud \vX_i\ud w_i w_i^2 K_h^2(\vx-\vX_i) f_p(w_i,\vX_i) - \left[\int\ud \vX_i\ud w_i w_i f_p(w_i,\vX_i)K_h(\vx-\vX_i)\right]^2 \right\}\nonumber
\\
&=& \frac{n\la w^2 \ra}{N^2}\int \ud \vx' K_h^2(\vx') - \frac{1}{N}\int \ud \vx \la \hat{f}\ra^2
= \frac{n\la w^2 \ra}{N^2 h^d} R(K) - \frac{1}{N}R(f) + \clo(h^2/N),
\eea
where $R(K)\equiv \int \ud \vy K^2(\vy)$. We can define an effective sample number
\be
\Nindep \equiv \frac{N^2}{\sum_i w_i^2} = \frac{(\sum_i w_i)^2}{\sum_i w_i^2}
\approx \frac{N^2}{n\la w^2\ra},
\label{eq:neff_indep}
\ee
so that the leading statistical variance scales $\propto 1/\Nindep$:
\be
\int \ud\vx \,\var[\hat{f}(\vx)] \approx \frac{1}{\Nindep h^d} R(K)  - \frac{1}{N}R(f) + \dots.
\label{covfhat}
\ee
For small $h$, the first term dominates.

The total mean integrated error of Eq.~\eqref{MISE} is the sum of the bias and statistical terms,
and evaluating the leading terms to get the Asymptotic Mean Integrated Squared Error ($\AMISE$) gives
\be
\AMISE = \int \ud \vx\, \var \hat{f} + \int \ud \vx \la\hat{f}-f\ra^2=
\frac{1}{\Nindep h^d} R(K)  - \frac{1}{N}R(f) + \frac{h^4}{4} \int \ud \vx \left([\text{Cov}(K)]^{ij} f^{(2)}_{ij}\right)^2
+ \dots,
\label{AMISE}
\ee

Minimizing this with respect to $h$ gives $h\propto [\Nindep] ^{-1/(4+d)}$, or explicitly an asymptotically-optimal kernel smoothing scale of
\be
h = \left( \frac{ R(K)d}{\Nindep I}\right)^{\frac{1}{4+d}},
\label{hopt}
\ee
where $I\equiv \int \ud \vx \left([\text{Cov}(K)]^{ij} f^{(2)}_{ij}\right)^2$.

Apart from the scaling with the effective number of samples $\Nindep$, $h$ also scales with the curvature of $f$ via the dependence on $I$: the larger the average squared second derivative, the more structure the gets smoothed out, and hence the smaller $h$ should be. But remember that this is specific to the simple linear kernel estimator of Eq.~\eqref{linearKDE}, assuming independent sample points.

\subsection{Multiplicative bias correction}
\label{sec:multbias}
Using boundary kernels renders estimates that are unbiased to $\clo(h^2)$. However, there is still a systematic broadening of peaks, which can lead to systematically overestimated errors unless there are sufficiently many samples that $h\ll 1$. We can do better (or save computing time by generating fewer samples), by using a higher-order estimator.

Note that the simple estimator is exactly unbiased if the density is flat (or linear). We can therefore try to \emph{flatten} the density before performing the convolutions. Specifically, doing the \emph{multiplicative bias correction} to form
\be
\hat{\hat{f}} = g (K_h \ast [H/g]),
\ee
where $g$ is an approximation to the shape of $f$, so that $H/g$ is nearly flat. Absent any prior information about the shape, the simplest thing to do is use $g = \hat{f}$, where $\hat{f}$ is a standard linear kernel density estimate; the $\hat{\hat{f}}$ estimator then has bias $\clo(h^4)$ away from boundaries (assuming sufficient smoothness of $f$)~\cite{Jones95}.
To improve the flattening near boundaries, we can take $\hat{f}$ to be the linear boundary kernel estimate from the Sect.~\ref{sec:boundary}.
In principle the flattening can also be iterated, but for good choice of smoothing widths usually little is to be gained (and iterations will not converge due to random fluctuations being magnified). The simple multiplicative bias correction method compares well with other higher-order kernel methods for many distributions~\cite{Jones97} and seems to work well in practice as long as the density is indeed sufficiently smooth. In principle different bandwidths can be used for the pilot estimator $g$ and the final estimate $\hat{f}$ (see e.g.~\cite{Hengartnera09} who recommend $g$ is over-smoothed compared to $\hat{f}$), but for simplicity we take them to be the same. Other approaches to bias reduction are possible, including the `data sharpening'~\cite{Choi99,Hall02} method, which is a special case of a more general diffusion approach~\cite{botev2010}.

Multiplicative bias correction produces smooth density estimates even with relatively small sample sizes. However, this smoothness comes with a caveat: it may mask sampling uncertainties that would be more visually apparent in lower-order estimates, where the lack of smoothness serves as a visual indicator of estimation uncertainty. \GetDist\ allows the multiplicative correction order to be changed as desired, but is set to first order by default (doing multiplicative bias correction once).

\subsection{Correlated samples}
\label{sec:correlations}

In reality, samples from MCMC are correlated. When expressed using weighted samples (where weights account for rejected proposal steps), both non-trivial weights and correlations exist between chain positions.  For a fixed number of samples, more correlation increases the uncertainty in our density estimates. 
An important and perhaps counterintuitive result is that correlations do not have a large effect on the optimal kernel bandwidth choice. This is because the main contribution of correlations to the variance is independent of the smoothing scale $h$: correlated errors between nearby points $\vx$ persist regardless of additional smoothing; see Refs.~\cite{Hall95,Skold03}.

We derived Eq.~\eqref{AMISE} for the kernel density error using $\Nindep$ independent weighted samples. With weights accounting for MCMC rejections, $\Nindep$  could be used as the effective sample number when doing MCMC sample bandwidth selection~\cite{Skold03}). However, this cannot be the full story when correlated samples are used with finite $h$ of practical interest. For example, proposals in orthogonal subdimensions could leave the parameter(s) of interest exactly unchanged between steps, even though they appear as different points in the full-dimensional parameter space. This could be remedied by using a parameter-dependent $\Nindep$ in Eq.\eqref{AMISE}, where the weights now count all consecutive identical points in the parameter space of the kernel density. However, it is also clear that very small changes in a parameter, for example due to accepted proposals along very nearly orthogonal eigendirections, should contribute nearly the same as exactly identical points. In other words, whether or not the correlation matters when determining the bandwidth depends on the shape of the correlation function; e.g., whether there is high probability for $|X_i-X_{i+k}| < h \sigma_X$, or whether the distribution is broad compared to $h\sigma_X$.

In detail, we have
\be
\int \ud\vx \, \hat{f}^2(\vx) = \frac{1}{N^2}\int \ud \vx \sum_{i,j}  w_i w_j K_h(\vx-\vX_i)
K_h(\vx-\vX_j)  = \frac{1}{N^2} \sum_{i,j} \ w_iw_j [K_h\ast K_h](\vX_i-\vX_j).
\ee
Assuming stationarity\footnote{Note that this is not valid for the output of nested sampling and other dynamic sampling methods; in these cases \GetDist\ currently simply treats the samples as independent, which could be improved in future.} leads to
\be
\int \ud\vx \, \la \hat{f}^2(\vx)\ra =
 \frac{n\la w^2 \ra}{N^2}\int \ud \vx' K_h^2(\vx')
 + \frac{2}{N^2} \sum_{k=1}^{n-1} (n-k) \la w_i w_{i+k} [K_h\ast K_h](\vX_i-\vX_{i+k})\ra,
\ee
and transforming $K_h\rightarrow K$ then gives
\be
\int \ud\vx \, \la \hat{f}^2(\vx)\ra =
 \frac{ R(K)}{\Nindep h^d}
 + \frac{2}{N^2 h^d} \sum_{k=1}^{n-1} (n-k) \left\la w_i w_{i+k} [K\ast K]\left(\frac{\vX_i-\vX_{i+k}}{h}\right)\right\ra.
 \label{varfcorr}
\ee
This makes it clear that the result depends on the number of sequences of points within distance $h$ of each other, as determined by the local $K\ast K$ filter. Note that the last term in Eq.~\eqref{varfcorr} contains a large contribution $\int \la \hat{f}(\vx)\ra^2$ from points that have close to the same value (a fraction $\clo(h/n)$ of the terms in the sum), even in the absence of correlations. If points separated by $k\alt \Delta$ are strongly correlated with $P(\vX_{i+k}|\vX_{i})\sim \delta(\vX_{i+k}-\vX_i)$, the second term also has a contribution $\sim R(K)\Delta/Nh^d$ that is the same order as the first; this limit is what is considered in Ref.~\cite{Skold03}, and accounts for rejection steps that leave the parameter value exactly unchanged.%
\footnote{
Note that if the (integer) weights are from chain rejections during MCMC (but we neglect correlations between accepted points), then $P(w) =(1-a)^{w-1}a$, where $a$ is the acceptance probability. Evaluating expectations gives
\be
\la w\ra = \frac{1}{a},
\qquad
\la w^2 \ra = \frac{2-a}{a^2}.
\ee
So for raw chains, neglecting correlations between accepted points, we have
\be
N^{\rm indep}_{\rm eff} \approx \frac{n  \la w \ra^2}{\la w^2\ra}
\approx \frac{N \la w \ra}{\la w^2\ra} \approx N \frac{a}{2-a}.
\ee
This relates results in terms of weights to results in the literature terms of acceptance probability (e.g. Ref.~\cite{Skold03}).
}

In general we can define a heuristic effective number of samples, explicitly dependent on which parameter subspace is included in the dimensions of $\vX$. We  estimate this from the samples as\footnote{It is often a good approximation to estimate the 2D result from the separate 1D results; in \GetDist\ there is an option whether to use the 2D expression or not ({\tt use\_effective\_samples\_2D}). Dependence of the optimal smoothing scale on $\NeffMCMC$ is quite weak, so a ballpark number is sufficient in most cases. A more optimal bandwidth estimator would not use a single $\NeffMCMC$, but account for anisotropy in the sampling statistics for sampling methods where different parameters are treated qualitatively differently or have different diffusion rates. }
\be
\NeffMCMC \equiv \frac{N^2}{
\sum_i w_i^2 +  2R(K)^{-1}\sum_k\sum_i \left(w_i w_{i+k} [K\ast K]([\vX_i-\vX_{i+k}]/h)- \hat{\mu}_K\right)},
\label{NeffCorr}
\ee
where the sum over $k$ can be taken only up to order of the correlation length ($\clo(L_X^s)$) where the terms are significantly non-zero (and hence is reasonably fast to evaluate), and $\mu_K \equiv \la w_i w_j [K\ast K]([\vX_i-\vX_{j}]/h)\ra$ takes out the $\sim \la \hat{f}\ra^2$ contribution expected for uncorrelated samples (estimated here roughly by a sum over widely separated small subset of samples). In the $h\rightarrow 0$ limit this definition therefore isolates the term that contributes to the total variance as
\be
\int \ud\vx \, \var \hat{f}(\vx) \approx \frac{1}{\NeffMCMC h^d} R(K) + \clo(1/N),
\ee
and hence includes the effect of exactly duplicated samples from MCMC rejection.
The definition of Eq.~\eqref{NeffCorr} obeys consistency under sample-splitting, so it does not matter how samples are grouped up in to weighted samples or split up, and for uncorrelated samples reduces to Eq.~\eqref{eq:neff_indep}.
More generally, Eq.~\eqref{NeffCorr} very roughly includes other tight short-range correlation effects from MCMC sampling (but also some additional covariance that is actually mostly $h$-independent, which ideally should not affect the bandwidth choice).
As defined $\NeffMCMC$ does however itself depend on $h$. We take a fiducial value $h\approx 0.2\sigma$ for estimating $\NeffMCMC$. Values of $\NeffMCMC$ typically lie between $\Nindep$ and the $\Neffvar$ defined in Eq.~\eqref{Neff} below that determines the sampling errors on parameter means.

\subsection{Choice of kernel bandwidth}

\begin{figure}
\begin{center}
\includegraphics[width=8.5cm]{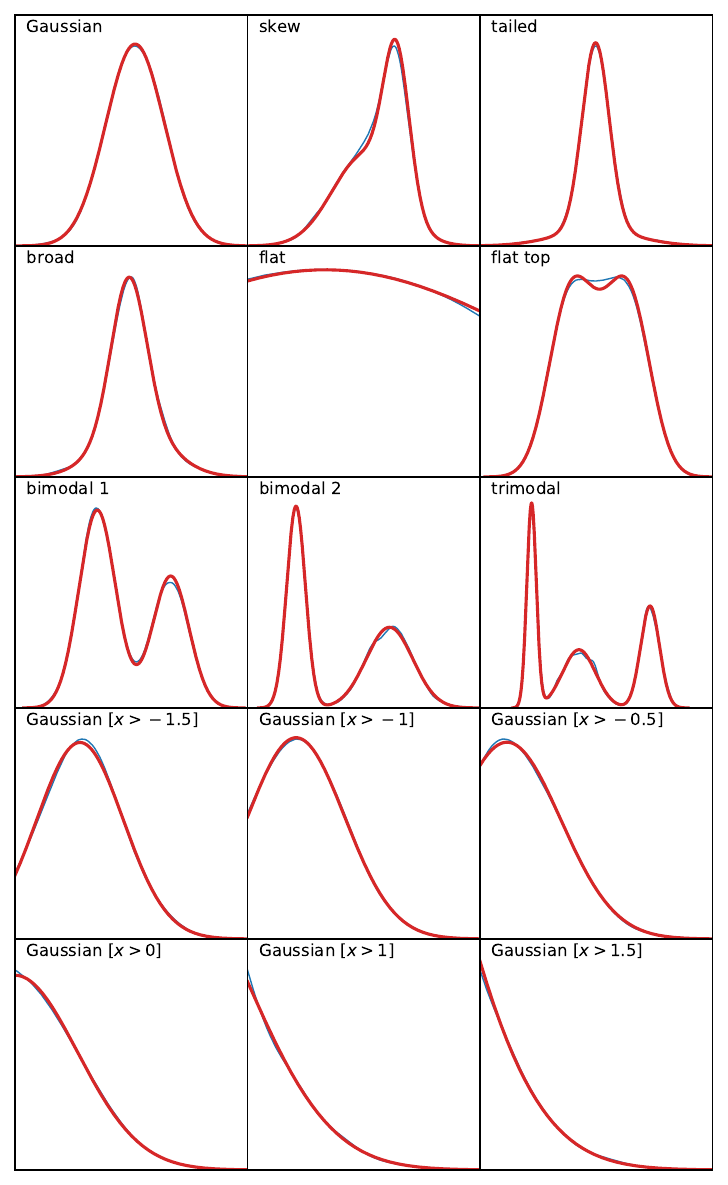}
\includegraphics[width=8.7cm]{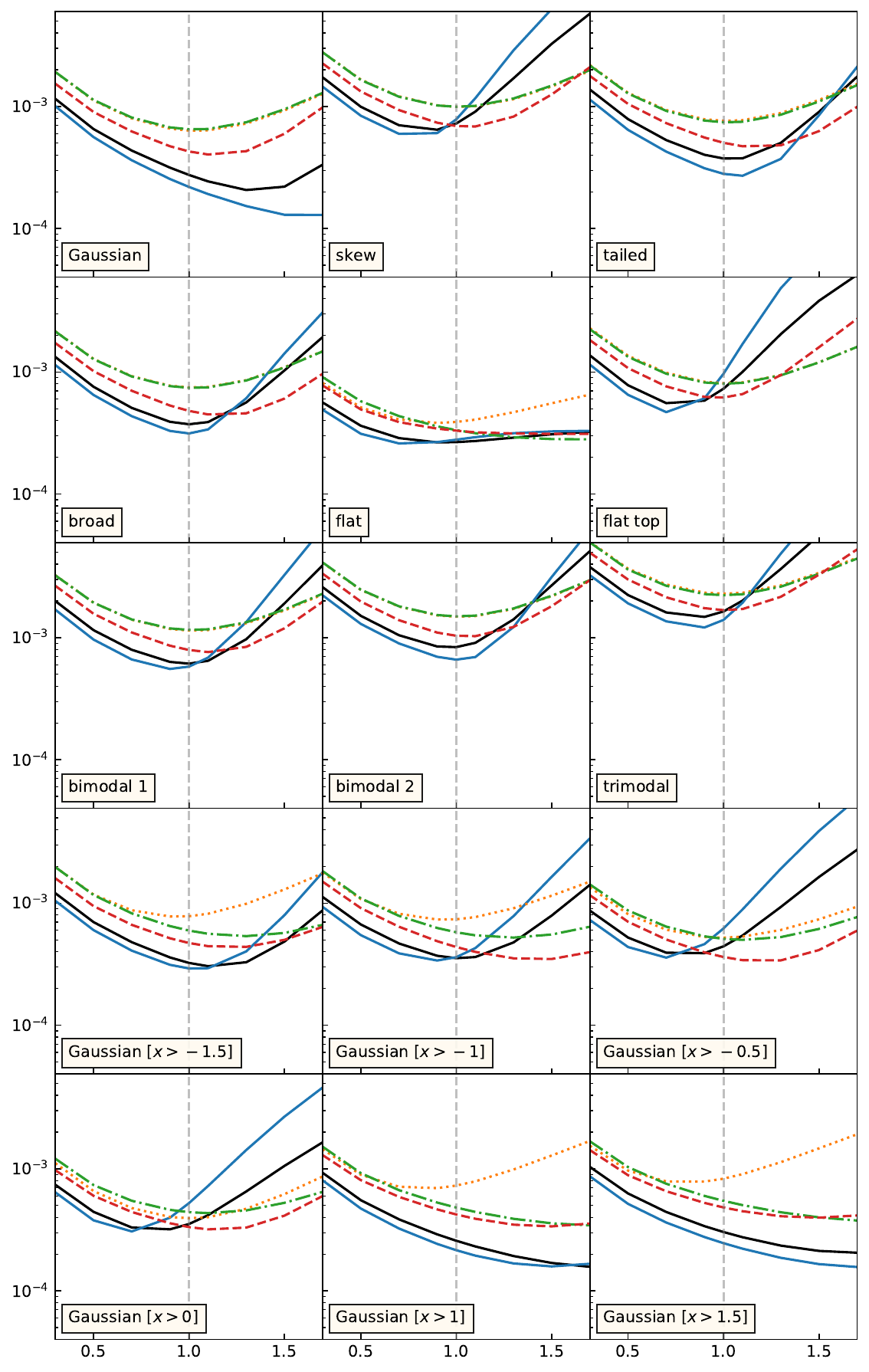}
\caption{\emph{Left}: a set of test Gaussian-mixture distributions, comparing the true distribution (red) with the density estimate using 10000 independent samples (blue) using multiplicative bias correction and a linear boundary kernel. The `Gaussian' panels at the bottom are truncated Gaussian distributions, and all distributions are normalized by the maximum value.
\emph{Right}: scaling of the average integrated squared error $\langle\int \ud x (\Delta f(x))^2\rangle/\int \ud x (f(x))^2$ of the density estimate, where the average is estimated using 1000 sets of 10000 samples for each distribution. The $x$-axis is a scaling relative to the automatically chosen kernel width (e.g. by Eq.~\eqref{h_MBC}), so that one corresponds to the performance with default settings. Lines compare different kernel estimates: solid lines use a multiplicative bias correction (MBC) and linear boundary kernel (black: default, blue: next-order multiplicative bias correction). Dotted is the basic Parzen kernel (for which the kernel-width estimator is optimizing), dot-dashed is with linear boundary correction, and dashed is using a second-order boundary-corrected kernel.
The MBC kernel width is suboptimally chosen for Gaussian, where the leading bias term happens to be zero, but about right in many other cases.
}
\label{tests1D}
\end{center}
\end{figure}

A good choice of kernel width is important to get good results: too broad, and features are washed out; too narrow, and sampling noise shows up. Recall from Eq.~\eqref{AMISE} that the Parzen–Rosenblatt estimator has bias $\clo(h^2)$, and the statistical variance goes as $\clo( [Nh]^{-1})$. Minimizing with respect to $h$ gave $h\propto N^{-1/5}$ (1D case of Eq.~\eqref{hopt}, corresponding to an overall convergence rate $\propto N^{-4/5}$). The constant in the optimal width depends on the distribution (and kernel). Assuming one-dimensional Gaussian distributions,  the rule of thumb for parameter $X$ is (`normal scale rule'):
\be
h = 1.06 \hat{\sigma}_X (\NeffMCMC)^{-1/5},
\label{ruleofthumb}
\ee
where $h$ is the standard deviation of the Gaussian smoothing Kernel to use, and $\hat{\sigma}_X$ is an estimate of the standard deviation of parameter $X$. In practice, for potentially non-Gaussian densities, $\hat{\sigma}_X$ can be set from a variety of scale measures. For example, a width based on central quantiles to avoid over-estimation due to broad tails, or a more refined method based on order statistics~\cite{Janssen95}. However, simple scale rules can be quite suboptimal for many non-Gaussian densities.
We only use a scale rule as a fallback when other methods fail and for choosing a fiducial scale for evaluating Eq.~\eqref{NeffCorr}.
To estimate $\hat{\sigma}_X$ we
follow a simplified version of Ref.~\cite{Janssen95} (taking $\hat{\sigma}_X = \text{min}[\sigma_X , R_{0.4}/1.048$], where $R_x$ is a the smallest parameter range enclosing $x$ of the probability ($\text{min} R_{0.4}=1.048$ for a unit Normal) and searching over ranges starting at $p=0, 0.1, 0.2...0.6$).

 An optimal bandwidth choice can be derived using Eq.~\eqref{hopt}. The only problem here is that the optimal bandwidth depends on second derivatives ($I$) of the (unknown) density $f$. Replacing the derivative term with an estimator gives so-called 'plug-in' methods, which can perform much better especially for multimodal distributions. For reviews and variations  of methods see
 e.g.~\cite{JonesMarronSheather96,eidous2010comparative,botev2010,heidenreich2013bandwidth}. The main problem is that to estimate the second derivative you need to use a bandwidth, which gives you a recursive unknown bandwidth problem.
  Ref.~\cite{botev2010} present a neat solution, where the optimal bandwidth is obtained as an equation fixed point that can be found numerically called the ``Improved Sheather-Jones" (ISJ) estimate. Using a Discrete Cosine Transform (DCT), this can also efficiently handle leading-order boundary effects along parameter axes, so that boundaries are not mistaken for large derivatives~\cite{botev2010}. The method only requires one DCT of the binned data and some binned array dot products, and hence is fast; we adopt it as our auto-bandwidth selector%
  \footnote{There can be multiple or no solutions to the fixed-point equation, esp. with some very flat bounded distributions. When there are multiple solutions we take the larger one, and when no solutions we use the fallback of Eq.~\eqref{ruleofthumb}.}.
  The DCT imposes even symmetry about boundaries, so we only use it for the bandwidth choice, not the actual KDE (the linear boundary kernel gives better accuracy by allowing general gradients at the boundaries).

With multiplicative bias correction the bias is higher order, with bias $\clo(h^4)$ away from boundaries, so the total error scales as $Ah^8 +B/(Nh)$. Optimization now gives $h\propto N^{-1/9}$ and overall convergence $\propto N^{-8/9}$. Again the proportionality constant will depend on the distributions, various examples are given in Ref.~\cite{Marzio09}. As a first guess we take the one-dimensional%
\footnote{In general we can replace $1/9$ with $1/(4p+1)$ for a higher order estimator where the leading bias goes as $h^{2p}$.} rule of thumb
\be
h =  h_{\rm ISJ} (\NeffMCMC)^{1/5-1/9}.
\label{h_MBC}
\ee
These smoothing widths are larger than for the basic Parzen–Rosenblatt estimator, and have lower statistical noise since the basic estimator is forced to have smaller widths to avoid significant bias. For $\NeffMCMC\sim \clo(1000)$, the smoothing width is about twice as broad as the basic estimator.
A more refined estimate could be made analogously to the ISJ method using the asymptotic error for the higher order method, but we have not attempted to implement this.  Eq.~\eqref{h_MBC} is somewhat too small for normal distribution (which happens to give zero leading bias for this estimator~\cite{Jones95}), but somewhat too large for some truncated Gaussian shapes. See Fig.~\ref{tests1D} for test results on various test distributions\footnote{The code describing the exact distributions and for reproducing the figures is at \url{https://github.com/cmbant/getdist/blob/master/getdist/tests/test_distributions.py},}. Higher-order bias correction can perform better, but starts to be more sensitive to having the bandwidth chosen optimally; as a default we use multiplicative boundary correction without iteration, which (in the test distributions) is almost always better than the Parzen estimator even when the auto-selected bandwidth is not optimal. In some cases using second order (once-iterated) multiplicative bias correction can give additional improvement. The \GetDist\ package has settings options to tune exactly which method is use if required.

\subsubsection*{Multivariate bandwidth matrix}

\begin{figure}
\begin{center}
\includegraphics[width=8.5cm]{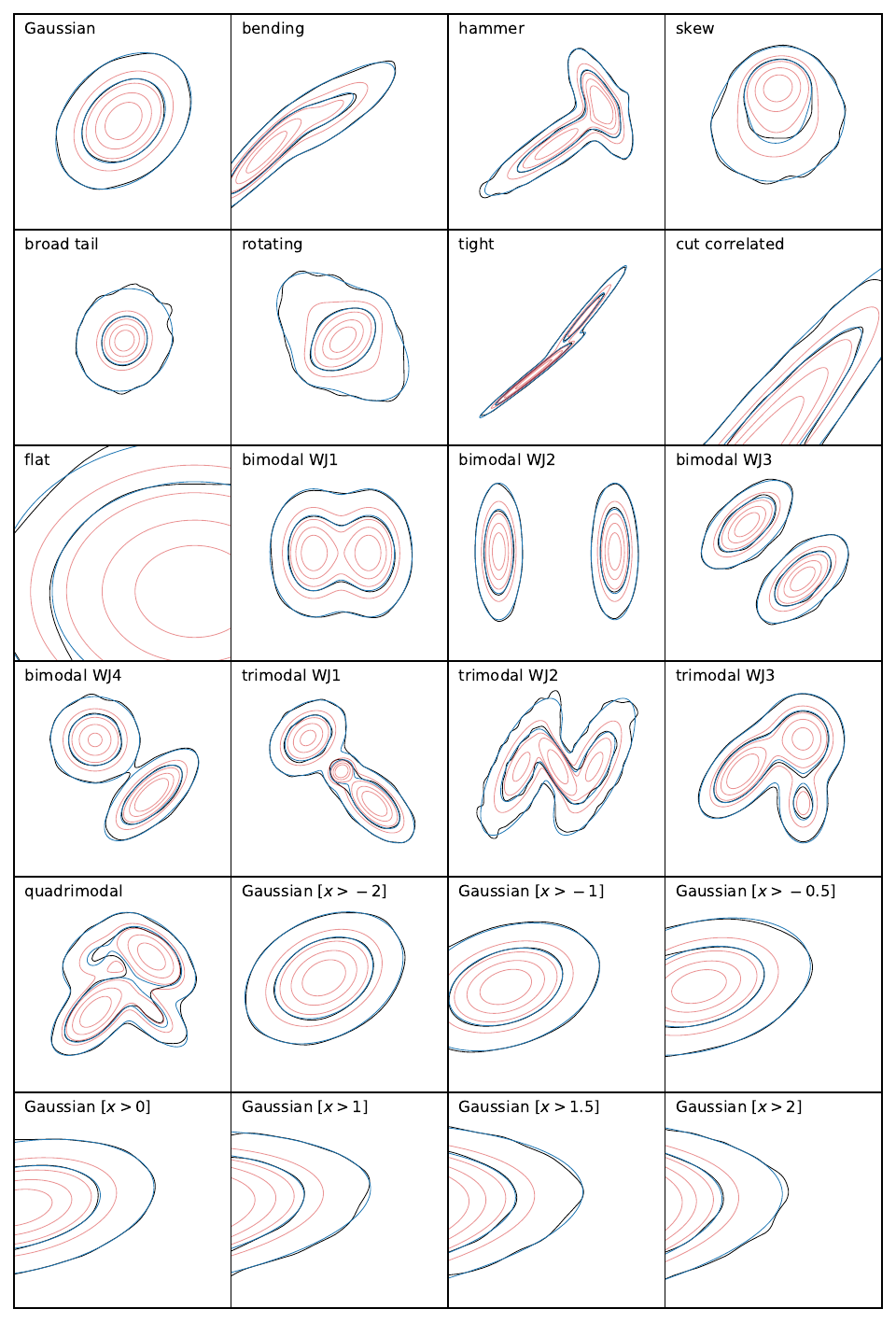}
\includegraphics[width=8.7cm]{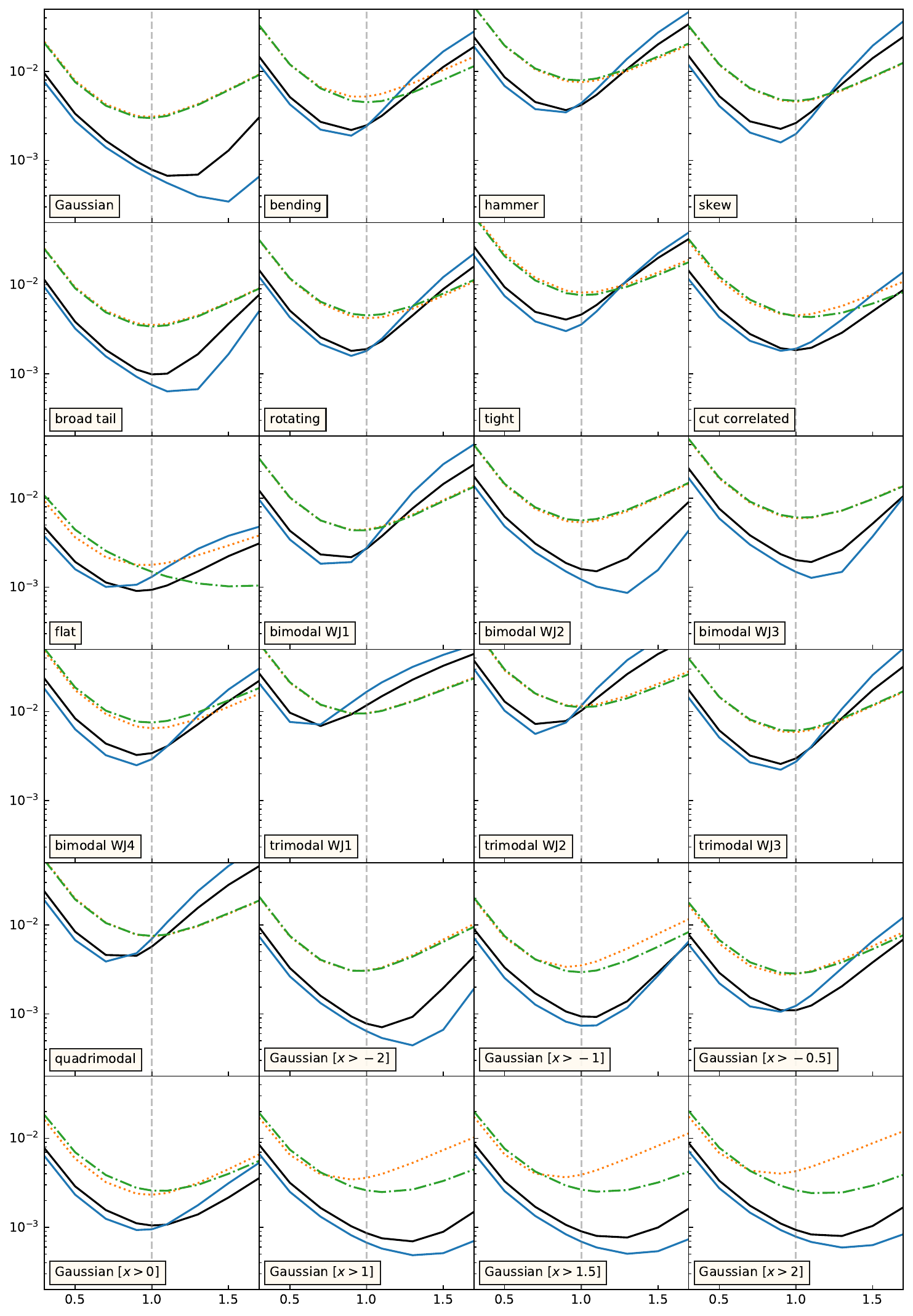}
\caption{\emph{Left}: A set of 2D Gaussian mixture distributions (WJ$x$ labels are from the same test distributions as Ref.~\cite{Wand93}), comparing true density contours (enclosing the 68\% and 95\% of the probability) with contours from density estimation using one set of 10000 samples. \emph{Right}: normalized average integrated squared error $\langle\int \ud x (\Delta f(x))^2\rangle/\int \ud x (f(x))^2$ from 500 simulations of 10000 samples, as Fig.~\ref{tests1D}.
In all but two of the trimodal examples the black lines (default is scale width one, with multiplicative bias correction and linear boundary kernel) give substantially lower error than the basic Parzen estimator (red dotted) and are more stable than the higher-order bias correction (blue).
}
\label{tests2D}
\end{center}
\end{figure}

For two-dimensional densities the optimal kernel will not in general be isotropic. In tightly-correlated distributions, the kernel shape should match the correlation direction to efficiently smooth along the degeneracy direction without causing spurious broadening in the well-constrained direction. In general, the shape could vary with position, but we assume the simplest case where the same kernel is used everywhere. This will work well in cases where there is only one clearly correlated direction, but may lead to sub-optimal results for more complex cases.
We define the kernel in terms of an isotropic Gaussian kernel $K(\vx)=K(|\vx|)$ and a kernel matrix
$\mM$ following e.g. Ref.~\cite{Wand93}. The non-isotropic kernel is then given by $K_M(\vx)=|\mM|^{-1/2}K(\mM^{-1/2}\vx)$, so that
\be
\int \ud\vx \, \var \hat{f}(\vx) \approx \frac{1}{\NeffMCMC}  |\mM|^{-1/2} \int \ud \vy K^2(\vy)+\dots \equiv  \frac{1}{\NeffMCMC}  |\mM|^{-1/2} R(K) + \dots.
\ee
If $K(\vx)$ has identity covariance, $\text{cov}(K) = \mI$, then $\mM$ is just the covariance of $K_M(\vx)$ and hence
\be
\AMISE \approx \frac{1}{\NeffMCMC} |\mM|^{-1/2} R(K)+ \frac{1}{4} \int \ud \vx \left( M^{ij} f^{(2)}_{ij}\right)^2
+ \dots
\label{multiAMISE}
\ee

If we parameterize the Gaussian kernel covariance as $\mM=\begm h_x^2 & ch_x h_y \\ ch_x h_y & h_y^2\enm$,
Eq.~\eqref{multiAMISE} becomes
\be
\AMISE\approx \frac{1}{4\NeffMCMC\pi h_x h_y \sqrt{1-c^2} } +
\frac{1}{4} \left[
h_x^4 \psi_{4,0} + h_y^4\psi_{0,4} + 2h_x^2 h_y^2(2c^2+1)\psi_{2,2} + 4c h_x h_y(h_x^2\psi_{3,1} + h_y^2\psi_{1,3})
\right],
\label{AMISE2Dfull}
\ee
where we defined $\psi_{m_1, m_2}$ as
\be
\psi_{m_1,m_2} \equiv
(-1)^{i+j}\int \ud\vx \left( \frac{\partial^{i+j}}{\partial{x_1^i}\partial{x_2^j}} f(\vx)\right)
\left(\frac{\partial^{p+q}}{\partial{x_1^p}\partial{x_2^q}} f(\vx)\right)
= \int \ud\vx f(\vx) \left(\frac{\partial^{p+q+i+j}}{\partial{x_1^{p+i}}\partial{x_2^{q+j}}} f(\vx)\right)
\ee
assuming no boundary terms, where $m_1=p+i$ and $m_2=q+j$ and $m_1+m_2$ is even. For $m_1$ and $m_2$ both even, $\psi_{m_1,m_2}$ (and corresponding bandwidths) can be estimated
following the fixed-point method\footnote{When there is no solution for the fixed point, we instead use a plugin estimate for the bandwidth used for estimating $\psi_{m_1,m_2}$.} of Ref.~\cite{botev2010}, where we assume an isotropic Gaussian kernel for evaluating $\psi_{m_1,m_2}$.
For the odd elements, the analogous argument to Ref.~\cite{botev2010} (Appendix E) using Eq. 3.2 from~\cite{wand1994multivariate} gives
an equate for the bandwidth for estimating $\psi_{m_1,m_2}$ as
\be
h_{m_1,m_2} = \left( \frac{8(1-2^{-m_1-m_2-1})}{3(\NeffMCMC)^2}\frac{\psi_{0,0} R(K^{(m_1,m_2)})}{ (\psi_{m_1, m_2+2} + \psi_{m_1+2, m_2})^2 }\right)^{1/(2m_1+2m_2+6)},
\label{diaghxhy}
\ee
where
\be
R(K^{(m_1,m_2)}) = \frac{(2m_1-1)!!(2m_2-1)!!}{2^{m_1+m_2+2}\pi}
\ee
and $\psi_{0,0}$ can be estimated using the method for even elements.

In the case that the correlation $c$ is zero, Eq.~\eqref{AMISE2Dfull} can be optimized analytically to give~\cite{wand1994multivariate}
\be
h_x = \left[ \frac{\psi_{0,4}^{3/4} R(K)}{\psi_{4,0}^{3/4}(\psi_{4,0}^{1/2}\psi_{0,4}^{1/2} + \psi_{2,2}) \NeffMCMC}\right]^{1/6}
\qquad h_y = ({\psi_{4,0}/\psi_{0,4}})^{1/4} h_x.
\label{hoptdiag}
\ee
In the general correlated case the minimum must be found numerically. In the specific case that the target distribution is Gaussian, the optimal Gaussian bandwidth matrix
covariance is~\cite{Wand93}
\be
\mM = \mC (\NeffMCMC)^{-1/3},
\ee
where $\mC$ is the sample covariance. This can be used to define a rule of thumb for Gaussian-like distributions, but in general (especially in the multimodal case) can be very bad.

There are several other issues here
\begin{itemize}
\item The bias term in Eq.~\eqref{AMISE2Dfull} is not guaranteed to be positive if $c\ne 0$, so numerical minimization can fail. (see Ref.~\cite{Duong03} for a possible alternative solution)
\item With boundaries, the even $\psi_{m_1,m_2}$ derivative terms can be approximated by imposing reflection boundary conditions (i.e. evaluating using DCT),
but with $m_1$ or $m_2$ odd, $\psi_{m_1,m_2}$ cannot be evaluated from the DCT transform (which assumes symmetry by construction).
They can be evaluated by FFT if there are no sharp boundaries, but there is no easy way to approximately account for boundaries in this case.
\item Since the  $\psi_{m_1,m_2}$  are evaluated using isotropic Gaussian kernels, they may be rather inaccurate if the optimal kernel is strongly elliptical.
\end{itemize}
We therefore adopt the following strategy:
\begin{itemize}
\item Assuming there are no boundaries, or a boundary in only one of the $x$ or $y$ directions (but not both), use the sample covariance to perform a Cholesky parameter rotation to define uncorrelated transformed variables.
  The Cholesky rotation is chosen so that if $x$ or $y$ has a boundary it remains unchanged, so the boundary
in the transformed parameters remains parallel to the edge of the DCT box. The transformed samples are scaled (so roughly isotropic) and binned, so that evaluation of $\psi_{m_1,m_2}$ using an isotropic kernel is not too suboptimal.
\item
If there is a boundary the even derivatives are evaluated following Ref.~\cite{botev2010} by DCT, and the optimal diagonal bandwidth matrix evaluated from Eq.~\eqref{hoptdiag}. This is then rotated back to the original coordinates.
\item If there are no boundaries, the even and odd $\psi_{m_1, m_2}$ derivatives are estimated, and~\eqref{AMISE2Dfull} is minimized numerically. If this fails, Eq.~\eqref{hoptdiag} is used as a fall back. The bandwidth matrix is then rotated back to the original coordinates.
\item If there are boundaries in both the $x$ and $y$ directions, a Cholesky rotation cannot preserve both boundaries, so the samples are not transformed.
The diagonal form of Eq.~\eqref{hoptdiag} is evaluated on the untransformed samples, unless the sample correlation is very high, in which case a Gaussian rule of thumb bandwidth is assumed using the sample covariance. When there are boundaries the fixed-point solution for the moment bandwidth can give
solutions that are substantially too large, in which case we fall back to a rule of thumb for the moment bandwidth.

\end{itemize}

The expected asymptotic scaling of the optimal bandwidths are $h\propto N^{-1/6}$ and $h\propto N^{-1/10}$ respectively for methods with quadratic and quartic bias.
With multiplicative bias correction we therefore scale the elements of the bandwidth matrix determined above to give
\begin{equation}
h_{x,y} = 1.1 h^{\rm ISJ}_{x,y} (N_{\rm eff})^{1/6-1/10},
\label{h_MBC2D}
\end{equation}
where the $1.1$ factor is empirically chosen.
(In general we can replace $1/10$ with $1/(2p+2)$ for a higher-order estimator where the leading bias goes as $h^{2p}$).
See Fig.~\ref{tests1D} for performance on typical distributions, showing that Eq.~\eqref{h_MBC2D} slightly underestimates the bandwidth for a Gaussian distribution (and tail-truncated Gaussians), but is a reasonable compromise for most other cases and gives significant performance gains compared to the basic Parzen estimator.

\section{Correlation lengths and sampling error on parameter means}

From MCMC, potentially with additional importance sampling, the samples generally have non-trivial weights and non-trivial correlations. Consider a sample estimate for the mean $\bar{X}$ of a parameter $X$, given by
\be
\hat{X} = \frac{1}{N} \sum_{i=1}^{n} w_i X_i.
\ee
From independent unit-weight samples, the variance of the mean estimator is $\sigma^2_X/N$; we can use this to define an effective $\Neffvar$ for the correlated weighted samples. The variance of $\hat{X}$ is given by
\be
\langle(\hat{X}-\bar{X})^2\rangle =
\frac{1}{N^2} \sum_{i=1}^{n}\sum_{j=1}^{n}  \langle w_i(X_i-\bar{X}) w_j(X_j-\bar{X})\rangle.
\ee
Defining $d_i\equiv w_i (X_i-\bar{X})$, for chains in equilibrium we should have
$\langle d_i d_j \rangle = C_d(|i-j|)$, where $C_d(k)$ is the autocorrelation function at lag $k$.
Using this
\bea
\langle(\hat{X}-\bar{X})^2\rangle &=& \frac{1}{N^2}\left[nC_d(0) + 2\sum_{k=1}^{n-1}(n-k) C_d(k)\right].
\eea
If we assume that the correlation length is much shorter than the chain length\footnote{Actually we don't need to do this, the finite estimator for the autocorrelation from the samples follows the original expression.}, so $k\ll n$ for terms which matter, this is
\be
\langle(\hat{X}-\bar{X})^2\rangle \approx \frac{n}{N^2}\left[C_d(0) + 2\sum_{k=1}^{\infty}C_d(k)\right].
\ee
We define this to be equal to $\sigma_X^2/\Neffvar$ so that
\be
\Neffvar \approx \frac{N^2\sigma^2_X}{n [C_d(0) + 2\sum_{k=1}^{\infty}C_d(k)]}
\label{Neff}
\ee

We can also define a correlation length by
\be
L^w_d \equiv \frac{n}{N\sigma_X^2} \left[C_d(0) + 2\sum_{k=1}^{\infty}C_d(k)\right],
\ee
so that $\Neffvar = N/L^w_X$.
For unweighted samples $L_X^w$ corresponds to the standard definition of the correlation length. For importance sampled chains, it is the length in `weight units' (it scales with the arbitrary normalization of the importance weights). We can also define a correlation length $L_X^s$ in `sample units', so that $\Neffvar = n/L^s_X$, which gives an idea of how independent the different points are.
In practice, to avoid sampling noise the upper limit for the sum is taken to be the lag at which the correlation has fallen to below some value (e.g. 0.05).

To estimate the error on Monte Carlo means, Eq.~\eqref{Neff} can be estimated quickly using weighted sample convolutions, and allows for both correlations and importance weights. In general there are correlations between parameters, so this is just an estimate for a single parameter, and will in general be optimistic (an upper limit).

\section{Credible intervals and contours}

Fully marginalized parameter constraints are often summarized as a mean and standard deviation (for distributions that are close to Gaussian), or a credible interval containing a given percentage of the posterior probability. For unimodal distributions these give a convenient reference for the where the bulk of the probability lies in parameter space. However, there is some freedom in exactly what quantities to report, and the \GetDist\ package has a number of setting parameters to determine exactly what is used for summary tables and figures. The defaults follow those used by the \planck\ cosmological parameter analysis as summarized in Ref.~\cite{Ade:2013zuv}.  

In addition to mean and variance, \GetDist\ will calculate $n$ credible intervals, by default three values set to $68\%$, $95\%$ and $99\%$. The calculation is a multi-step process designed to robustly handle various distribution shapes, including those affected by parameter boundaries:
\begin{enumerate}
  \item \textbf{Initial Parameter Range Estimation:}
  For each parameter, an initial working range (\code{range_min}, \code{range_max}) is determined from the weighted samples. This range excludes extreme outliers by spanning from the \code{range_confidence} quantile to the quantile of total weight $1-\code{range_confidence}$. By default $\code{range_confidence}=0.001$, so the
      range includes 99.8\% of the probability. 
      
    \item \textbf{Incorporate Prior Boundaries:}
    Specified hard prior boundaries (\code{limmin}, \code{limmax}, often from sampler metadata) are considered:
    \begin{itemize}
        \item If a prior boundary is close to the initial sample range (from step 1), the corresponding \code{range_min} or \code{range_max} is adjusted to this prior boundary, and a flag (\code{has_limits_bot} or \code{has_limits_top}) is set to indicate an active prior at that end.
        \item If a prior boundary is well outside the initial sample range, it is ignored for the purpose of this range setting.
        \item If no prior boundary is active at an end (either unspecified, or ignored because it's too far), the range at that end is slightly extended. This helps ensure the subsequent density estimation captures the tails.
    \end{itemize}
    The scale used to determine closeness and extension is based on an estimate of the distribution's characteristic scale.

    \item \textbf{1D Kernel Density Estimation (KDE):}
    A 1D kernel density estimate (KDE) of the marginalized posterior for the parameter is computed over the \code{range_min} to \code{range_max} established in steps 1 and 2. This KDE accounts for boundary effects from 
    any active priors and is normalized so its peak value is one.
    
    \item \textbf{Define ``Significant Density" Threshold:}
    For each desired confidence level, calculate a threshold density ratio \\ \code{max_frac_twotail}. By default this taken to be the ratio of the probability density at the tails of a Gaussian distribution (at the points defining the specified confidence level, e.g., approximately $\pm 1\sigma$ for 68\%) to its peak density. This threshold is used to determine if the density at the edge of a range is sufficiently small for a tail limit to be meaningful. For $68\%$ and $95\%$ limits, \code{max_frac_twotail} is approximately $0.6099$ and $0.1465$ respectively.

    \item \textbf{Assess Density at Range Edges:}
    The KDE density is evaluated at \code{range_min} and \code{range_max}. 
    \\Flags (\code{marge_limits_bot}, \code{marge_limits_top}) are set to indicate if the distribution appears significantly truncated by a boundary prior; e.g. \code{marge_limits_bot} is true if there is a hard prior at that end (\code{has_limits_bot} is true) \textit{and} the KDE density at that end is greater than \code{max_frac_twotail} (from step 4). 

    \item \textbf{Handle Fully Prior-Dominated Cases:}
    If both \code{marge_limits_bot} and \code{marge_limits_top} are true (i.e., the distribution has high density up against active hard priors at both ends, like a uniform posterior filling its prior range), no interval limits are reported for this confidence level, as the parameter is effectively constrained by these priors rather than forming distinct tails.

    \item \textbf{Compute credible interval from KDE:}
    
        If step 6 does not apply, the algorithm computes a highest density interval containing the required fraction of the total probability using the following procedure:
 \begin{itemize}
        \item The KDE grid points are sorted in descending order of density values
        \item The cumulative sum of these sorted density values (each weighted by the grid spacing) is calculated
        \item For a confidence level $p$ (e.g., 0.68), the algorithm finds the density threshold where the cumulative sum equals $(1-p)$ times the total probability
        \item The algorithm then identifies the outermost points where this density threshold intersects with the original KDE curve, using linear interpolation between grid points
        \item These intersection points define the limits (\code{tail_limit_bot}, \code{tail_limit_top}) of the credible interval
    \end{itemize}
    This approach identifies a density threshold and finds the outermost points where the density equals this threshold. For unimodal distributions, this ensures all points within the interval have higher probability density than any point outside it. However, for multimodal distributions, the interval may include lower-density regions between modes, as the algorithm reports the outermost limits that encompass all regions above the threshold.

    \item \textbf{Report One-Tailed Limits:}
    If, after step 7, one of \code{marge_limits_bot} or \code{marge_limits_top} is true (indicating the posterior has significant density up against an active prior at one boundary) while the other is false (indicating the posterior falls off towards the other end of the range), a one-tailed limit is reported. This limit is derived directly from the sorted weighted samples to ensure robustness.
    \begin{itemize}
        \item For an X\% confidence level (e.g., 95\%): If the upper tail is small, an upper limit is reported, defined as the value below which X\% of the total sample weight lies (the X-th percentile).
        \item Conversely, if the lower tail is small, a lower limit is reported, defined as the value above which X\% of the sample weight lies (the $(100-X)$-th percentile).
    \end{itemize}

    \item \textbf{Report Two-Tailed Limits:}
    Otherwise, if both \code{marge_limits_bot} and \code{marge_limits_top} are false (meaning the posterior density is low at both ends of the established range relative to \code{max_frac_twotail}), a two-tailed interval is reported
    \begin{itemize}
   \item Calculate an equal-tailed two-tail limit (\code{tail_confid_bot}, \code{tail_confid_top}) directly from the samples. 
   \item Calculate the KDE densities at these two values
   \item If the absolute difference between these densities is less than a fraction \code{credible_interval_threshold} (default 0.05) of the peak KDE density, the sample-based (\code{tail_confid_bot}, \code{tail_confid_top}) equal-tailed interval is reported. This is preferred for its numerical stability when the densities at the tails are similar.       
     \item  Otherwise, report the \code{tail_limit_bot}, \code{tail_limit_top} credible interval.
       This ensures the reported interval limits have a similar posterior density.
   \end{itemize}
\end{enumerate}

This procedure intelligently determines the most appropriate way to report parameter constraints. For symmetric distributions, it uses equal-tailed intervals derived directly from sample weights, which provide numerical stability. For asymmetric distributions, it switches to highest-density intervals where the posterior density is equal at both limits, avoiding the misleading inclusion of low-probability regions that can occur with equal-tailed intervals. The algorithm properly handles boundary cases, reporting one-tailed limits when appropriate (such as when a parameter is only bounded from one side) and correctly accounting for prior-truncated posteriors. For multimodal distributions, the reported interval spans from the leftmost to the rightmost high-density regions, which may include lower-density regions between modes. When generating LaTeX output, \GetDist\ applies heuristics to select an appropriate number of significant figures for reporting means and credible intervals, ready for publication use.

Credible regions are also used for 2D plots. 
These are found from the 2D KDE by identifying density contours (iso-probability density lines) such that the integral of the KDE over the area enclosed by a contour corresponds to the required probability fraction (e.g., 68\%, 95\%). For non-unimodal or complex 2D distributions, these regions may be disconnected.

\section{Discussion}

While \GetDist\ demonstrates robust performance for many common applications, there are several important assumptions and limitations to consider. Future development could focus on several key areas:

\begin{enumerate}
    \item Kernel estimator optimization for non-stationary sampling distributions (e.g., nested sampling results). The current implementation assumes stationarity in the sampling process, which may not hold for all sampling methods.
    
    \item Integration of general prior boundary effects into kernel optimization. The current approach to kernel smoothing scale selection does not explicitly account for prior boundaries except for the case where priors are aligned with the parameter axes, potentially leading to suboptimal bandwidth choices in more general cases.
    
    \item Handling of complex likelihood topologies, particularly for highly multimodal distributions with varying characteristic scales across different modes. While the current methods remain functional in these cases, the use of a global smoothing kernel may be far from optimal. Significant improvements could be achieved by implementing more sophisticated approaches that can adapt to local distribution characteristics, e.g. using clustering methods~\cite{meszaros2024}.
    
    \item Optimization of kernel widths for specific computational tasks. The current implementation optimizes kernel widths solely for density estimation. For other applications, such as calculating tail confidence limits, different optimization criteria may be more appropriate.
\end{enumerate}

\GetDist\ can easily be used with any numpy array of samples, but also has built-in support for samples from the {\sc Cobaya} sampling package~\cite{Torrado:2020dgo}, automatically propagating sample names, labels and prior bounds, and well as an import function for \texttt{ArviZ}~\cite{arviz_2019} (as used for example by \texttt{PyMC}). It also supports a general text-based file format for samples, e.g. as used by the \planck\ analysis and output by some other samplers, but tighter integration with other sampling packages may be possible.

The code base is maintained as an open-source project on GitHub, and welcomes community contributions.

\section{Acknowledgements}
I thank Jesus Torrado for work on the Cobaya interface, and Jesus and other github users for contributions.
I acknowledge support from the European Research Council under the European Union's Seventh Framework Programme (FP/2007-2013) / ERC Grant Agreement No. [616170] and support by the UK STFC grants ST/P000525/1 and ST/X001040/1.

\providecommand{\aj}{Astron. J. }\providecommand{\apj}{ApJ
  }\providecommand{\apjl}{ApJ
  }\providecommand{\mnras}{MNRAS}\providecommand{\prl}{PRL}\providecommand{\prd}{PRD}\providecommand{\jcap}{JCAP}\providecommand{\aap}{A\&A}

\end{document}